\documentclass[11pt,twoside]{article}

\usepackage{asp2006}
\usepackage{epsf}
\usepackage{lscape}

\begin{document}
\title{Topics on bar and bulge formation and evolution}   
\author{Athanassoula E.}   
\affil{LAM, OAMP, 2 place Le Verrier, 13248 Marseille cedex 04,
  France}

\begin{abstract} 
I discuss results from the COSMOS survey, showing that the fraction of
disc galaxies that is barred decreases considerably with look-back time
from $z\sim0.2$ to $z\sim0.8$. This decrease is more important for small mass
and low luminosity spirals. Classical bar formation theory provides a
promising framework for understanding these results.

I also discuss the formation of discy bulges using $N$-body simulations
reproducing well the properties of observed discy bulges. Thus, these
simulated discy bulges have the shape of a disc, they have S\'ersic
profiles with small values of the shape index and their size is of the
order of a kpc. They are formed by radial inflow of material driven by
the bar and are thus composed of both gas and stars and have a
considerable fraction of young stars. They can harbour spiral
structure, or an inner bar.
\end{abstract}


\section{Introduction} 

I will discuss two specific topics on bar and bulge formation and
evolution. The first one concerns the formation of bars in time, 
measured from a large sample of disc galaxies observed with COSMOS. My
collaborators for this work are 
Kartik Sheth, Debra Meloy Elmegreen, Bruce Elmegreen, Peter Capak, 
Roberto Abraham. Richard Ellis, Bahram Mobasher, Mara Salvato,
Eva Schinnerer, Nick Scoville, Lori Spalsbury, Linda Strubbe, Marcella Carollo,
Michael Rich and Andrew West. The second one discusses results of
$N$-body simulations describing the formation of discy bulges. My
collaborators for this work are Clayton Heller and Isaac Shlosman. 

\section{Evolution of the bar fraction in COSMOS: Quantifying the
  assembly of the Hubble sequence}     

Bars drive the angular momentum exchange between the various
components of disc galaxies and therefore drive their evolution.
When did they form? Where all bars formed at the same time, or not?
Did specific types of barred galaxies form their bars before others?
To answer questions such as the above, my collaborators and I analysed
the fraction of disc galaxies that are barred in a large sample of
galaxies from the COSMOS 2-square degree field
(\citealp{Scoville07a,Scoville07b}). Setting thresholds for
brightness, photometric accuracy, redshift, type and inclination angle,
we obtain a sample of 2157 luminous, far from edge-on, spiral
galaxies. Thus, our sample is an order of magnitude larger and is
based on substantially deeper imaging data than other samples used in
previous bar fraction investigations
(\citealp{abraham99,sheth03,elmegreen04,jogee04}).  
 
Each galaxy in our sample was analysed for the existence of a bar with
two different methods and the results were cross-checked. For the first
method we fitted ellipses on the isophotes of the galaxy images and
identified bars from dual criteria on the profiles of the ellipticity
and of the position angle of the ellipses. We also identified bars
visually. 

\begin{figure}
\plotone{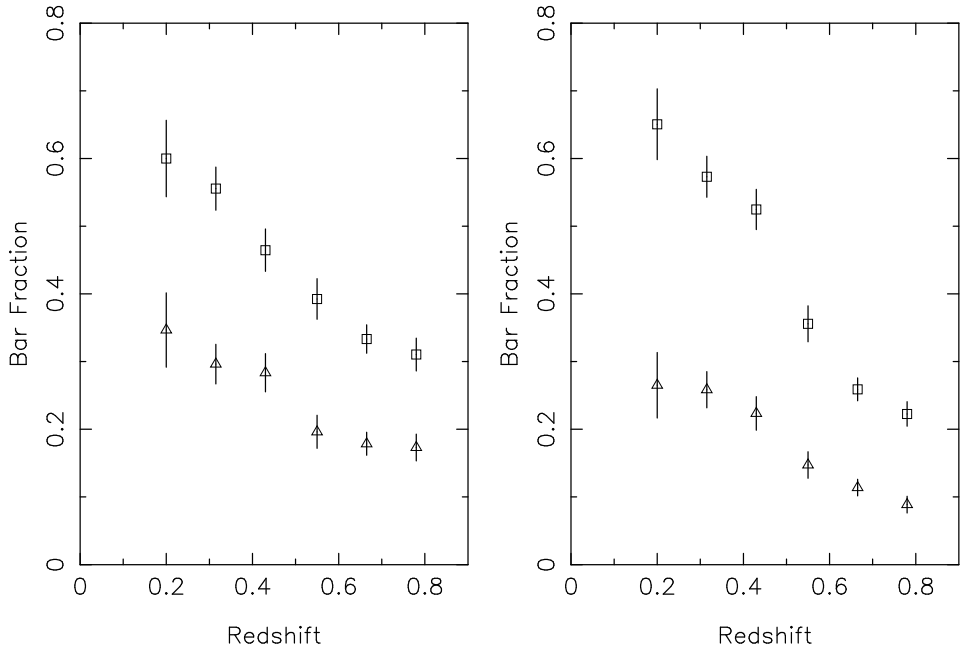}
\caption{Fraction of disc galaxies that are barred (open squares), or
  strongly barred (open diamonds) as a function of redshift. The left
  panel gives results from the visual classification, while the right
  panel gives results from the classification based on the ellipse
  fits (see text). The error bars reflect the statistical uncertainty 
  and are obtained from the expression
  $\left(f\left[1-f\right]/N\right)^{1/2}$, where $f$ is the fraction
  and $N$ is the number of galaxies.}
\label{fig:cosmos}
\end{figure}

From this analysis we find that the fraction of disc galaxies which
is barred is not constant with time, but increases strongly with
decreasing redshift from $z$ = 0.84 to $z$ = 0.2. The size of our
sample allows us to use six time bins and still have sufficient galaxies
in each bin to reach statistically safe results. These results can
be visualised in Fig.~\ref{fig:cosmos}, where I show both the
total fraction of bars and the fraction of strong bars. I also show
separately the results from the visual classification (left panel) and
the classification from ellipse fitting (right panel). Although the
thresholds for strong bars and the threshold between barred and
non-barred galaxies may differ somewhat between the two
classifications, it is evident that the increase of the bar fraction
with decreasing redshift is equally clear in both cases. 

We also find that the bar fraction in spiral 
galaxies depends on the stellar mass. Thus, the bar fraction in very
massive, luminous spirals is about constant out to $z~\sim$ 0.84,
whereas for the low mass spirals it declines significantly with
increasing redshift beyond $z$ = 0.3 (see Figures 2 and 3 in Sheth et
al. 2007). This result is a signature of downsizing 
and is intimately connected with what we may call the dynamical
maturity of discs. 

The increase in the bar fraction with decreasing redshift from $z$ = 0.84 to
$z$ = 0.2 can be understood within the framework of classical bar 
formation theory. $N$-body simulations have long suggested that bars
form spontaneously in galactic discs, usually on relatively short
dynamical timescales. There are, however, two ways of slowing this
down. One is to increase the halo mass fraction within the disc
radius, and the other is to heat up the disc
(\citealp{athasell86,ath02,ath03}). Although in many ways very different,
both these effects allow to slow
down the formation of the bar. Thus, the time it takes for an unbarred disc
galaxy to become barred can 
vary widely. In cold, disc-dominated cases, the bar forms within a
Gyr or less. Sufficiently hot discs embedded in very massive halos
can stay unbarred several Gyrs. Such a delay might well explain the
time evolution of barred galaxy fraction shown in
Fig.~\ref{fig:cosmos}. It could also explain the downsizing signature
found here, for two reasons. Observations
show that the halo-to-disc mass ratio is higher in low mass, low
luminosity galaxies than in bright, massive galaxies (\citealp{Bosma04,
KranzSR03}) so that bars are expected to grow later in
the former, as we indeed find here. Furthermore, there are suggestions
that the former are dynamically hotter than the latter \citep{Kassin+07}.
Although this picture could be complicated by interactions, or
eventually by bar dissolution, it 
can, nevertheless, provide a framework within which our results can be
understood.  

More discussion and analysis of these results can be found in Sheth et
al. (2007).

\section{Formation of discy bulges}    

Bulges are not a homogeneous class of objects, to a large extent due
to the different definitions used so far. \cite{ath05}
distinguished three different types of bulges : Classical, which
resemble ellipticals in many ways; boxy/peanut bulges, which are just
parts of bars seen near edge-on; and discy bulges, which are given the name
`bulges' because of their contributions to the inner parts of the
radial luminosity profiles.

\begin{figure*}
\plotfiddle{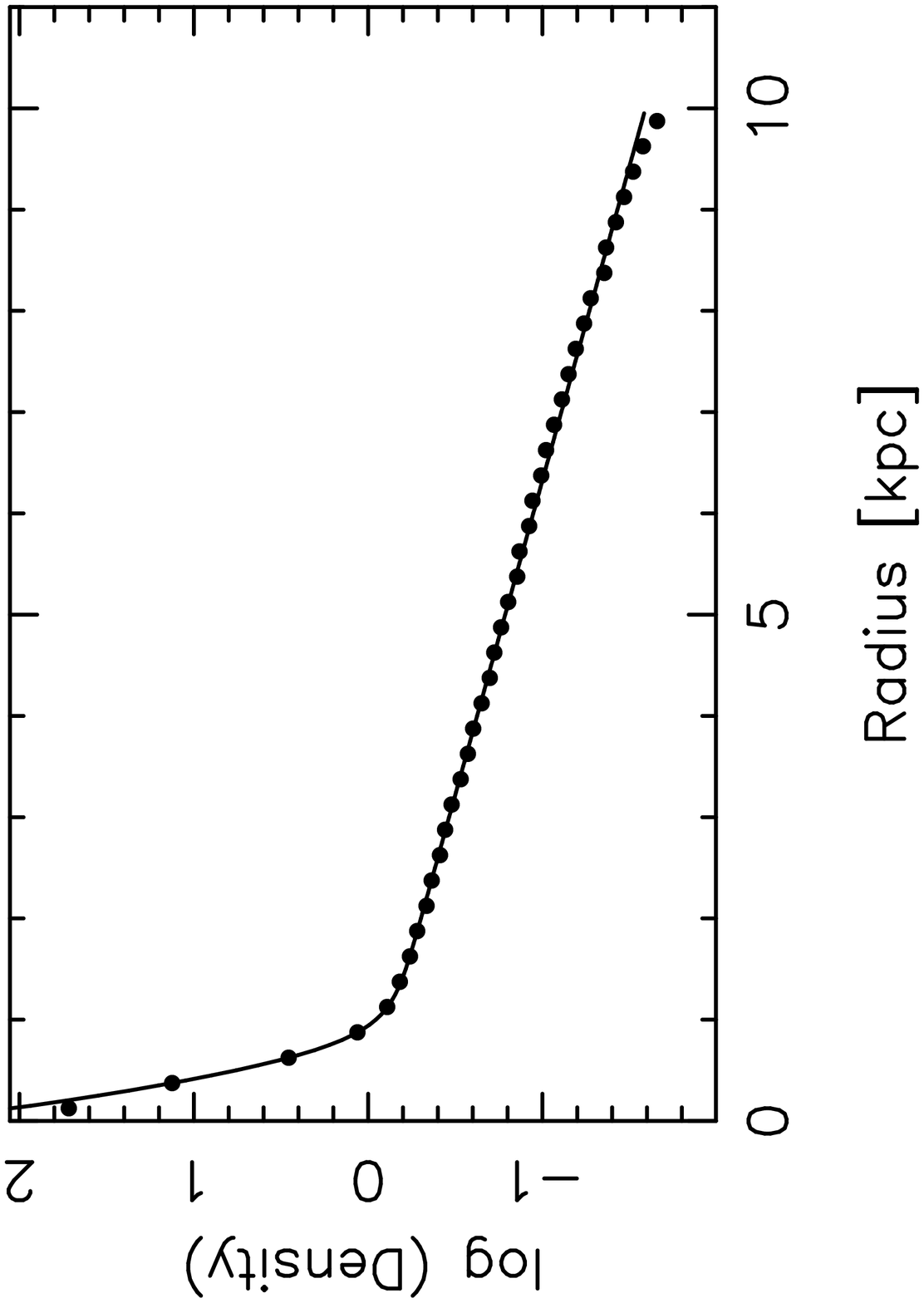}{0cm}{-90}{36}{36}{-50}{45}
\label{fig:densitydiscy}
\end{figure*}

I present here results from simulations similar to those of
\citet*{HellerSA07a} and \citeyearpar{HellerSA07b}, i.e. simulations
including gas, stars and dark matter, as well as star formation,
cooling and feedback.  
Several non-axisymmetric components -- such as a triaxial halo, an oval disc,
an inner and/or an outer bar -- form during these simulations. Their
interactions 
give very interesting dynamical phenomena (Heller et al. 2007a, b),
while they induce considerable inflow
and gaseous high density inner discs. This high gas concentration in
the central  
area triggers considerable star formation, resulting in a disc-like
central, high-density object, which, seen face-on, is often
somewhat oval. It has many properties similar to
those of discy-bulges. For example, it has, in many cases,
sub-structures, like an inner bar. Furthermore,  
a decomposition of the stellar radial density profiles 
gives results in good agreement with observations. An example of such a
radial projected surface 
density profile is given in Fig.~\ref{fig:densitydiscy}, together
with a fit by an exponential disc and a S\'ersic component. Note that

\begin{minipage}{3.5cm}
{\small {Figure 2. Radial
  projected stellar density profile in arbitrary units. Radii are measured in
  kpc. The dots give the
  simulation results and the solid line the fit by an exponential
  disc and a S\'ersic component. 
}}

\vspace{1.5cm}
\end{minipage}

\noindent
the fit is excellent, all the way to the outer parts of the disc,
roughly at 10 kpc. In this example, the disc 
scale-length is $\sim$2.7
kpc, i.e. very realistic, while the S\'ersic index is
$\sim$1, in good agreement with observed discy bulges (see
\cite{KormendyKennicutt04} for a review).

\acknowledgements I thank the Agence Nationale de la Recherche for
grant ANR-06-BLAN-0172.
The HST COSMOS Treasury program was supported through NASA grant
HST-GO- 09822. 


\end{document}